\documentclass[floatfix,aps,prb,twocolumn,noshowpacs,preprintnumbers,amsmath, 
amssymb]{revtex4} 
\usepackage{epsfig} 
\usepackage{graphicx}% Include figure files 
\usepackage{dcolumn}% Align table columns on decimal point 
\usepackage{bm}% bold math 
 
\newcommand {\E}[1]{\text{exp} \left[ #1 \right]} 
 
\begin {document} 
\title {Phase slips in superconducting films with 
constrictions} 
\author {Sang L. Chu, A. T. Bollinger, and A. Bezryadin} 
\affiliation {Department of Physics, University of Illinois at Urbana-Champaign, 
Urbana, IL 61801-3080} 
\date {\today} 
 
%%%%%%%%%%%% 
% ABSTRACT % 
%%%%%%%%%%%% 
\begin {abstract} 
A system of two coplanar superconducting films seamlessly 
connected by a bridge is studied. We observe two distinct 
resistive transitions as the temperature is reduced. The first 
one, occurring in the films, shows some properties of the 
Berezinskii-Kosterlitz-Thouless (BKT) transition. The second 
apparent transition (which is in fact a crossover) is related to 
freezing out of thermally activated phase slips (TAPS) localized 
on the bridge. We also propose a powerful indirect experimental 
method allowing an extraction of the sample's zero-bias resistance 
from high-current-bias measurements. Using direct and indirect 
measurements, we determined the resistance $R(T)$ of the bridges 
within a range of {\em eleven orders of magnitude}. Over such 
broad range, the resistance follows a simple relation $R(T)=R_N 
\text{ exp } [-(c/t)(1-t)^{3/2}]$, where $c=\Delta F(0) / kT_c$ is 
the normalized free energy of a phase slip at zero temperature, 
$t=T/T_c$ is normalized temperature, and $R_N$ is the normal 
resistance of the bridge. 
\end {abstract} 
\maketitle 
 
%%%%%%%%%%%%%%%% 
% INTRODUCTION % 
%%%%%%%%%%%%%%%% 
\section {Introduction} 
Thermally activated vortex-like excitations (topological defects) 
of the superconducting condensate is the primary source of 
dissipation in mesoscopic superconducting structures.\cite 
{tinkham_text} These fluctuations take different forms in 
one-dimensional (1D) and two-dimensional (2D) systems. In 2D thin 
films the fluctuations are known to be broken vortex-antivortex 
pairs \cite {berezinskii, kosterlitz_thouless, kosterlitz, 
beasley_etal, halperin_nelson, minnhagen, bancel_gray, 
hebard_fiory, rosario_etal, strachan_etal} while in 1D wires the 
resistance is due to phase slips.\cite {little, langer_ambegaokar, 
mccumber_halperin, lukens_etal, newbower_etal, sharifi_etal, 
lau_etal, tinkham_lau, rogachev_bezryadin, bollinger_etal} One 
important difference between these two types of fluctuations is 
that vortices and antivortices form bound pairs below a certain 
critical temperature, known as the Berezinskii-Kosterlitz-Thouless 
(BKT) transition temperature, while phase slips and 
anti-phase-slips are unbound at any finite temperature. Thus the 
resistance of 1D wires is greater than zero at any finite 
temperature due to the presence of phase slips, which are 
described by the theory of Langer, Ambegaokar, McCumber, and 
Halperin (LAMH), \cite{langer_ambegaokar, mccumber_halperin} 
 
Here we report a study of structures in which both types of 
fluctuations can coexist, namely thin films containing 
constrictions, which are comparable in size to the coherence 
length. The goals of this work are (i) to test the applicability 
of the LAMH theory for short and rather wide constrictions and 
(ii) to test the effect of the vortex-antivortex sea existing in 
the thin film banks adjacent to the constriction on the phase 
slippage rate on the constriction itself. For this purpose we 
fabricate and measure a series of thin superconducting MoGe 
films\cite{graybeal_thesis} (which are about 15 $\mu$m wide) 
interrupted by constrictions or ``bridges'' (see Fig. 
\ref{fig:sample}a). The width of the narrowest point of the 
bridges is in the range of 13-28 nm, i.e. a few times larger than 
the coherence length (we estimate $\xi(0)\approx 7$ nm for our 
MoGe films\cite{bezryadin_etal}). Two resistive transitions are 
observed in such samples indicating that vortex-antivortex pair 
binding-unbinding transition (if any) and thermally activated phase slip 
processes occur separately.  For $T>T_{\text{BKT}}$ the 
contribution of vortex-antivortex pairs is dominant. On the other 
hand, below $T_{\text{BKT}}$ the transport properties are 
determined by the phase slip process on the bridge, which may be 
regarded as a vortex-antivortex pair breaking assisted by the 
bridge.  
 
Using direct and indirect techniques we have tracked the 
sample's resistance within a range of eleven orders of magnitude. 
The resulting $R(T)$ curves are compared with the LAMH theory. 
Regardless of the large width of the bridges and their shortness, 
the shape of the measured $R(T)$ curves is in perfect agreement 
with the overall shape of the curves computed using the standard 
LAMH theory (note that this theory was originally derived for very 
long wires that are much thinner than the coherence length). The 
only disagreement found with LAMH is that the pre-exponential 
factor had to be modified in order to obtain a reasonably low 
critical temperature of the bridges. (The critical temperature is 
used as an adjustable parameter in the fitting procedure.) 
Following the argument of Little\cite{little} we arrive at the 
conclusion that the pre-exponential factor should be simply $R_N$ 
and obtain a good agreement with measured curves. The measurements 
show that the bridges with intermediate dimensions (i) allow phase 
slippage which does not quench at any finite temperature, (ii) 
behave independently of the thin film banks, and (iii) exhibit a 
higher rate of phase slippage in the cases when the width of the 
bridge is smaller and therefore when the coupling between the thin 
film banks is weaker. 
 
Before presenting our experimental results we give a brief summary 
of the BKT theory of topological phase transitions and the LAMH 
theory of thermally activated phase slips. In thin superconducting 
films, even in the absence of a magnetic field, an equal 
population of free vortices and antivortices is expected to 
occur. The BKT theory predicts a universal jump in the film 
superfluid density $n_s$ at the characteristic temperature 
$T_{\text{BKT}}$, lower than the mean field critical temperature 
of the film $T_{c0}$. Such a jump is related to the 
vortex-antivortex pair binding through a logarithmic interaction 
potential between free vortices \cite {kosterlitz_thouless, 
goldman_etal}. Applied currents can break bound pairs producing 
free vortices and leading to non-linear $V(I)$ curves. Above 
$T_{\text{BKT}}$ the linear resistance of a film is given by the 
Halperin-Nelson (HN) formula\cite {hebard_fiory, halperin_nelson} 
\begin {equation} 
\label {eqn:hn} R_{\text{HN}} = 10.8 b R_{n, \text{f}} \ \E {-2 
\sqrt{b (T_{c0} - T_{\text{BKT}}) / (T - T_{\text{BKT}})}} 
\end {equation} 
where $R_{n, \text{f}}$ is the normal state resistance per square 
of the film and $b$ is a non-universal constant. Note that the HN 
equation predicts zero resistance for temperatures below the BKT 
phase transition temperature $T_{\text{BKT}}$. 
 
The LAMH theory\cite {tinkham_text,little,langer_ambegaokar, 
mccumber_halperin} applies to narrow superconducting channels, in 
which thermal fluctuations can cause phase slips, i.e. jumps by $2 
\pi$ of the phase difference of the superconducting order 
parameter. In unbiased samples the number of phase slips (which 
change the phase difference by $2 \pi$) equals the number of 
anti-phase-slips (which change the phase difference by $-2 \pi$). 
An applied bias current pushes the system away from the 
equilibrium and the number of phase slips becomes larger than the 
number of anti-phase-slips. Thus a net voltage appears on the 
sample, which can be calculated, following LAMH, as 
$V=\hbar\dot{\phi}/2e$ (below we will also discuss an alternative 
approach to the voltage definition). Here $\hbar$ is Planck's 
constant, $e$ is the electron charge, and $\dot{\phi}$ is the rate 
of change of the phase difference between the ends of the wire. 
During the phase slip process the energy of the system increases 
since the order parameter becomes suppressed to zero in the center 
of the phase slip. Thermal activations of the system over this 
free energy barrier $\Delta F(T)$ occur at a rate given by 
$(\Omega(T)/2\pi)e^{-\Delta F / kT}$. If the bias current is not 
zero, then the net rate of the phase slippage is 
$\dot{\phi}=\Omega (T)(e^{-\Delta F_{+}(I) / kT}-e^{-\Delta 
F_{-}(I) / kT})$. Here $I$ is the bias current, and $\Delta F_{+}$ 
and $\Delta F_{-}$ are the barriers for phase slips and 
anti-phase-slips correspondingly (these two barriers become equal
to each other at zero bias current). 
The attempt frequency derived from a time-dependent 
Ginzburg-Landau (GL) theory, for the case of a long and thin wire, 
is\cite {mccumber_halperin} 
\begin {equation} 
\label{eqn:attempt_frequency} \Omega (T) = \frac{L}{\xi(T)} 
\frac{1}{\tau_{\text{GL}}} \left( \frac{\Delta F}{kT} 
\right)^{1/2} 
\end {equation} 
where $T$ is the temperature of the wire, and $L/\xi(T)$ is the 
length of the wire measured in units of the GL coherence length 
$\xi(T)$. The attempt frequency is inversely proportional to the 
relaxation time $\tau_{\text{GL}} = \pi \hbar / 8k(T_c -T)$ of the 
time-dependent GL theory, with $T_c$ being the mean field critical 
temperature of the wire (or of the bridge, as in our discussions 
below). The factor $\left( \Delta F / kT \right)^{1/2}$ 
provides a correction for the overlap of fluctuations at different 
places of the wire and the factor $L/\xi(T)$ gives the number of 
statistically independent regions in the 
wire\cite{mccumber_halperin}. The free energy barrier for a single 
phase slip is given \cite{langer_ambegaokar, tinkham_lau} by 
\begin {equation} 
\Delta F = \frac{8 \sqrt{2}}{3} \frac{H_{c}^2 (T)}{8 \pi} A \xi 
(T), 
\end {equation} 
which is essentially the condensation energy density $H_c^2 
(T) / 8 \pi$ multiplied by the effective volume  
$8 \sqrt{2} A \xi (T) / 3$ of a phase slip ($A$ is the 
cross-section area of the wire). 
 
A bias current $I$ causes a non-zero voltage (time averaged) given 
by 
\begin {equation} 
\label {eqn:lamh_voltage} V = \frac{\hbar\Omega (T)}{e} e^{-\Delta 
F / kT} \sinh (I/I_0) 
\end {equation} 
where $I_0 = 4ekT/h$ ($I_0 = 13.3$ nA at $T=1$ K). Differentiation 
of this expression with respect to the bias current $I$ gives the 
differential resistance 
\begin {equation} 
\label {eqn:lamh_dv_di} dV/dI = \frac{\hbar\Omega (T)}{eI_0} 
e^{-\Delta F / kT} \cosh (I/I_0) 
\end {equation} 
The dependence of the attempt frequency and free energy on the bias current is 
neglected in this derivation. In the limit of low currents $I \ll 
I_0$, Ohm's law is recovered 
\begin {equation} 
\label {eqn:lamh_ohmic} R_{\text{LAMH}} (T) = \frac{\hbar\Omega 
(T)}{eI_0} e^{-\Delta F / kT} = R_q \left( \frac{\hbar\Omega 
(T)}{kT} \right) e^{-\Delta F / kT} 
\end {equation} 
where $R_q=h/(2e)^2=6.5$ k$\Omega$. In this approach the 
fluctuation resistance does not have any explicit dependence on 
the normal resistance of the wire. 
 
%%%%%%%%%%%%%%%%%%%%%%%%%%%%%%%% 
% FABRICATION AND MEASUREMENTS % 
%%%%%%%%%%%%%%%%%%%%%%%%%%%%%%%% 
\section {Experimental Setup} 
\begin {figure} [h] 
\begin {center} 
\epsfig{file=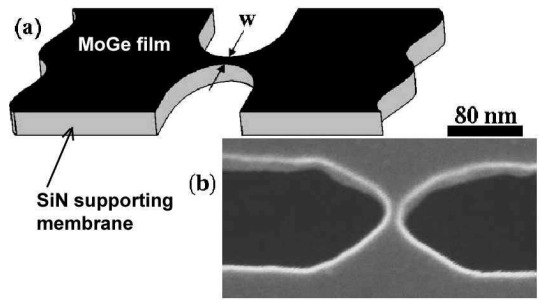, width=2.8 in} \caption {(a) Sample schematic. 
MoGe film (black) of thickness $d=$ 2.5--4.5 nm is deposited over 
a SiN membrane (gray) substrate with a constriction of width $w$. 
(b) An SEM micrograph of a typical sample. The MoGe coated SiN 
bridge (gray) is suspended over a deep trench (black).} \label 
{fig:sample} 
\end {center} 
\end {figure}

%We turn now to the experiment. 
The sample geometry is shown schematically in Fig. 
\ref{fig:sample}a.  The fabrication is performed starting with a 
Si wafer covered with SiO$_2$ and SiN films. A suspended SiN 
bridge is formed using electron beam lithography, reactive ion 
etching, and HF wet etching.\cite {bezryadinJVST} The bridge and 
the entire substrate are then sputter-coated with amorphous 
Mo$_{79}$Ge$_{21}$ superconducting alloy, topped with a $2$ nm 
overlayer of Si for protection\cite{sputter}. The resulting 
bridges are $100$ nm long with a minimum width $w \approx $ 13-28 
nm as measured with a scanning electron microscope (SEM) (Fig. 
\ref{fig:sample}b). All samples are listed in Table I. 
 
Transport measurements are performed in a pumped $^4$He cryostat 
equipped with a set of rf-filtered leads. The linear resistance 
$R(T)$ is determined from the low-bias slope (the bias current is 
in the range  of 1-10 nA) of the voltage versus current curves. 
The high-bias differential resistance is measured using an ac 
excitation on top of a dc current offset generated by a 
low-distortion function generator (SRS-DS360) connected in series 
with a 1 M$\Omega $ resistor.  One sample was measured down to the 
m$\Omega$ level using a low temperature transformer manufactured 
by Cambridge Magnetic Refrigeration. 
 
\begin {table} [b] 
\begin {center} 
\begin {tabular}{c c c c c c c} 
\hline \hline Sample & \ $w$ (nm) & \ $R_N$ ($\Omega$) & \ $T_{c}$ 
(K) & \ $T_{c0}$ (K) & 
\ $d$ (nm) & \ $\beta$ \\ 
\hline 
A1 & $27 \pm 4$ & $1380$ & $3.88$ & $3.90$ & $2.5$ & $1.47$\\ 
B1 & $13 \pm 4$ & $1650$ & $4.80$ & $4.91$ & $3.5$ & $0.723$\\ 
B2 & $28 \pm 4$ & $1320$ & $4.81$ & $4.91$ & $3.5$ & $1.10$\\ 
C1 & $13 \pm 4$ & $1440$ & $5.16$ & $5.50$ & $4.5$ & $0.653$\\ 
C2 & $27 \pm 4$ & $680$ & $5.39$ & $5.50$ & $4.5$ & $2.21$\\ 
\hline \hline 
\end {tabular} 
\end {center} 
\caption {\label {tab:samples} Sample parameters, including the 
width of the constriction ($w$), determined from SEM images, 
normal resistance of the bridge ($R_{N}$), determined from the 
$R(T)$ curves (at a temperature slightly below the resistive 
transition of the thin film banks), critical temperature 
($T_{c}$), determined from $R_{\text{WL}}(T)$ fits given by Eq. 
\eqref{eqn:modified_lamh}), critical temperature of the film 
($T_{c0}$), film thickness ($d$), and a geometrical fitting 
parameter ($\beta$).} 
\end {table}

%%%%%%%%%%%%%%%%%%%%%%%% 
% RESULTS % 
%%%%%%%%%%%%%%%%%%%%%%%% 
\section {Results} 
 
\begin {figure} [t] 
\begin {center} 
\epsfig{file=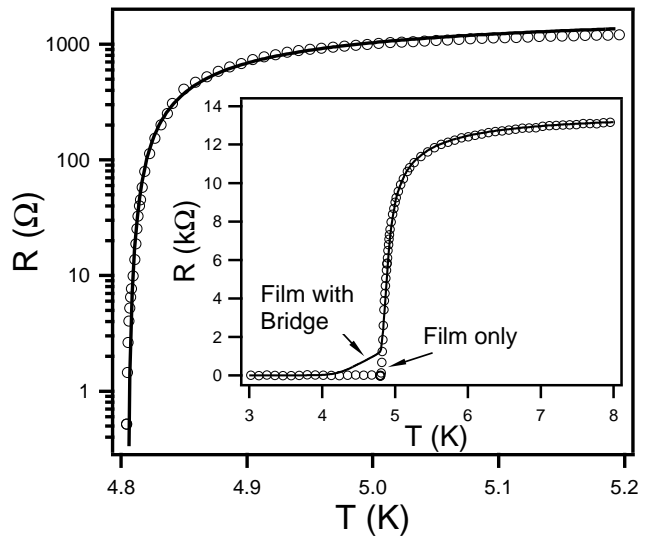, width=3.3 in, height=2.78in} 
\caption {Low-bias resistance versus temperature dependence (open 
circles), measured on a thin film ($d = 3.5$ nm) without 
constriction. The solid line is a fit to the Halperin-Nelson 
theory (Eq. \eqref{eqn:hn}). (Inset): Resistance of the film 
without constriction (multiplied by a constant factor), shown as 
open circles, is compared to the sample with a hyperbolic bridge 
(B2), shown by the solid line. The only qualitative difference is 
the presence of a ``resistive tail'', observed on all samples with 
constrictions.} \label {fig:film+bridge} 
\end {center} 
\end {figure}

First we compare a sample with a hyperbolic constriction 
(``bridge sample'') with a reference sample, which is a plain MoGe 
film of the same thickness, without any constriction (``film 
sample''). Both are fabricated on the same substrate simultaneously. 
A resistive transition measured on the film sample is shown in 
Fig. \ref{fig:film+bridge}.  The HN fit generated by Eq. 
\eqref{eqn:hn} is shown as a solid line and exhibits a good 
agreement with the data, yielding a 
BKT transition temperature of $T_{\text{BKT}} = 4.8$ K and the mean 
field critical temperature $T_{c0} = 4.91$ K.  Such good fit suggests that the 
transition observed in the banks might be the BKT transition, although a 
more extensive set of experiments is necessary in order to prove this 
assumption rigorously.  As expected, 
$T_{\text{BKT}}$ is slightly lower than $T_{c0}$. The inset of 
Fig. \ref{fig:film+bridge} compares the $R(T)$ measurements of the 
``film'' (open circles) and the ``bridge'' (solid line) samples. 
At $T = 4.8$ K the $R(T)$ curve for the film sample crosses the $R 
= 0$ axis with a nonzero (and large) slope, in agreement with the  
behavior predicted by the HN resistance equation \eqref{eqn:hn}.  
Nevertheless, unlike the film sample, the 
bridge sample shows a non-zero resistance even below the BKT 
transition temperature predicted by Eq. \eqref{eqn:hn}.  
Such resistive tails, occurring at $T<T_{\text{BKT}}$, 
have been found in all samples with constrictions.

In Fig. \ref {fig:lamh_many_samples} the $R(T)$ curves for five 
samples with bridges are plotted in a log-linear format. The 
resistance of sample B1 has been measured down to the m$\Omega$ 
range using a low temperature transformer. Two resistive 
transitions are seen in each curve as the temperature decreases. 
The first transition is the superconducting transition in the 
thin film banks adjacent to the bridge.  The second transition 
corresponds to the resistive tail mentioned above. In order to 
understand the origin of the second transition it should be 
compared to the LAMH theory. 
 
%%%%%%%%%%%%%%%%%%%%%%%% 
%DISCUSSION% 
%%%%%%%%%%%%%%%%%%%%%%%% 
\section {Discussion} 
 
Below we analyze the resistive tails found on samples with 
constrictions and demonstrate that they are caused by the phase 
slip events localized on the bridges and behave independently of 
the adjacent thin film banks. The analysis indicates that no BKT 
(no vortex-antivortex binding within the constrictions) 
or any other type of transition occurs on the constrictions and 
that the phase slips and anti-phase-slips are unpaired at any 
nonzero temperature due to thermal fluctuations. This is 
demonstrated below by fitting the $R(T)$ curves with the 
LAMH-like fitting curves. 
 
\subsection {LAMH attempt frequency for a short bridge} 
In order to compare our results to the LAMH we have to take into 
account the small length of the bridge, which does not allow more 
than one phase slip at a give time. Therefore the attempt 
frequency $\Omega(T)$ of Eq. \eqref{eqn:attempt_frequency} can be 
simplified. First, it has a term $L/\xi(T)$ that accounts for the 
number of independent sites where a phase slip can occur \cite 
{mccumber_halperin}. Since each of our samples has only one 
narrow region where phase slip events can happen, we take 
$L/\xi(T) = 1$.  Second, the coefficient $(\Delta F /kT)^{1/2}$ 
which takes into account possible overlaps of phase slips at 
different places along the wire\cite {mccumber_halperin} is 
taken to be unity also. This is because for short hyperbolic 
bridges (not much longer than the coherence length) it is 
reasonable to expect that there is only one spot, i.e. the 
narrowest point of the bridge, where phase slips occur. As a 
result, we obtain the attempt frequency for a short hyperbolic bridge 
$\Omega_{\text{WL}}=1/\tau_{\text{GL}}$ (the abbreviation ``WL'' 
stands for ``weak link''). This attempt frequency can be combined 
with the usual form of the LAMH resistance in Eq.  
\eqref{eqn:lamh_ohmic} and can be used to fit 
the experimental $R(T)$ curves (below the resistive transition of 
the films). Although such fits  
follow the data very well, there is one inconsistency that is 
they require the critical temperature of the 
bridge to be chosen higher than the critical temperature of the 
films, which is unphysical for such system. We attempt to modify the 
pre-exponential factor in order to resolve this inconsistency, as 
discussed below. 
 
\begin {figure} [h] 
\begin {center} 
\epsfig{file=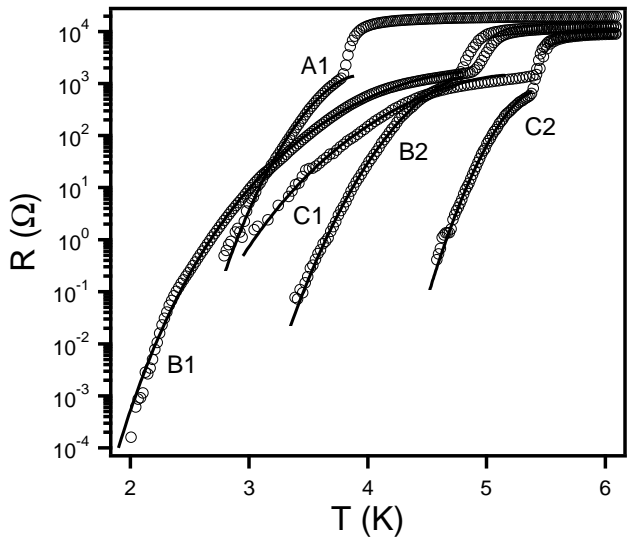, width=3.5in} \caption {Low-bias resistance 
for five different samples with bridges. The parameters of the 
samples are given in Table \ref{tab:samples}. The data points are 
shown by open symbols. Solid lines are fits to the ``bridge'' phase 
slip model given by Eqs. \eqref{eqn:modified_lamh0} and 
\eqref{eqn:modified_lamh}.} \label {fig:lamh_many_samples} 
\end {center} 
\end {figure} 
 
\subsection {Modification of the prefactor} 
 
Since the exact expression is unknown, we approximate the 
resistance of a constriction (weak link) as 
\begin {equation} 
\label {eqn:modified_lamh0} R_{\text{WL}}(T) = R_N\ e^{-\Delta 
F_{\text{WL}}/kT}. 
\end {equation} 
The exponential factor here is that of the LAMH theory and the 
prefactor is simply the normal resistance of the bridge.  This 
expression (Eq. \eqref{eqn:modified_lamh0}) can be justified by 
the following argument: the duration of a single phase slip (i.e. 
the time it takes for the order parameter to recover) is 
$\sim\tau_{\text{GL}}$ and the number of phase slips occurring per 
second is $\sim\Omega_{\text{WL}}(T) \ \E{-\Delta F(T)/kT}$, with 
the attempt frequency being the inverse GL relaxation time 
$\Omega_{\text{WL}}=1/\tau_{\text{GL}}$, as was argued above. 
Therefore the time fraction during which the constriction is 
experiencing a phase slip (i.e. when superconductivity is 
suppressed on the bridge) is the product of these two values, i.e. 
$f=(\tau_{\text{GL}})(1/\tau_{\text{GL}}) \ \E{-\Delta F(T)/kT} = 
\E{-\Delta F(T)/kT}$. Following Little,\cite{little} it can be 
assumed that the bridge has the normal resistance $R_N$ during the 
time when a phase slip is present (i.e. when the bridge is in the 
normal state), and the resistance is zero otherwise (when there is 
no phase slip). Thus we arrive at the averaged resistance for a 
bridge or a small size weak link $R_{WL}=f \times R_N +(1-f)\times 
0 = R_N \ \E{-\Delta F/kT}$ as in Eq. \eqref{eqn:modified_lamh0}. 
Note that unlike in the LAMH theory, in the present formulation the 
fluctuation resistance is directly linked to to the normal state 
resistance of the sample. 
 
In order to compare Eq. \eqref{eqn:modified_lamh0} to the 
experimental results, an explicit expression for the energy 
barrier $\Delta F_{\text{WL}}$ for a phase slip localized on the bridge 
is required. Starting with the usual 
form\cite{tinkham_lau} derived for a long 1D wire and some well 
known results from BCS and GL theory,\cite 
{tinkham_text,tinkham_lau} we find that $\Delta F_{\text{WL}} 
(0)=(8\sqrt2/3)(H_c^2(0)/8\pi)A\xi(0) = 0.83kT_cR_qL/R_N \xi(0)$ 
where $L$ is the length of the wire. Using $R_N=\rho_nL/A$, the 
free energy barrier for a weak link is 
\begin {equation} 
\Delta F_{\text{WL}} (0) = 0.83 k T_c \frac{\beta w d R_q}{\rho_n \xi(0)} 
%\frac{8\sqrt2}{3} \frac{H_c^2(0)}{8 \pi} \beta wd \xi(0) = 
\end {equation} 
where $w$ is the width of the bridge, $d$ is the film thickness, 
$\rho_n$ is the normal resistivity, and $A=wd$. The parameter 
$\beta$ measures the ratio of the phase slip length along the 
bridge to the effective length of a phase slip in a 1D wire, which 
is equal to $8 \sqrt{2} \xi (T) / 3$. Finally, assuming the 
same temperature dependence of the barrier as in the LAMH theory, 
i.e. $\Delta F(T)= \Delta F(0)(1-T/T_c)^{3/2}$, we arrive at the 
expression for the bridge fluctuation-induced resistance: 
\begin {equation} 
\label {eqn:modified_lamh} R_{\text{WL}}(T) = R_N\ \E{-0.83 
\frac{\beta w d R_q}{\rho_n \xi(0)} \left( 1 - \frac{T}{T_c} 
\right)^{3/2} \frac{T_c}{T}}. 
\end {equation} 
 
The fits generated by Eq. \eqref{eqn:modified_lamh} are shown in 
Fig. \ref{fig:lamh_many_samples} as solid lines.  An impressively 
good agreement is found for all five samples. In particular, 
sample B1 measured using the low-temperature transformer, shows an 
agreement with the predicted resistance $R_{\text{WL}}$ over about 
seven orders of magnitude, down to a temperature that is more than
two times lower compared to the critical temperature of the sample.
Only two fitting parameters are used: 
$\beta$ and $T_c$ (listed in Table \ref{tab:samples}). The other 
parameters required in Eq. \eqref{eqn:modified_lamh}, including 
$R_N$, $d$, $w$, $\xi (0)$, and $\rho_n\approx 180$ $\mu \Omega$ 
cm are known.\cite {bezryadin_etal, tinkham_lau, lau_etal} The 
fits give quite reasonable values for the critical temperature of 
the bridges, in the sense that they are slightly lower than the 
corresponding critical temperatures of thin films of the same 
thickness, as expected. This fact supports the validity of Eq. 
\eqref{eqn:modified_lamh0}.   Such good agreement also indicates 
that the dissipation in a thin film with a constriction at 
$T<T_{\text{BKT}}$ is solely due to thermal activation of phase 
slips on constrictions.  As expected, $\beta \approx 1$ for all 
samples and the larger $\beta$ values are found on wider 
constrictions. 
 
\begin {figure} [t] 
\begin {center} 
\epsfig{file=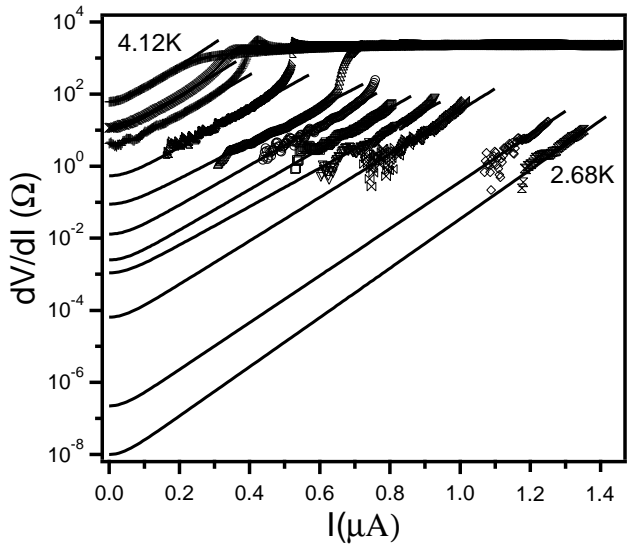, width=3.5in} 
\end {center} 
\caption {Differential resistance as a function of the dc bias 
current for sample B2.  Experimental data are denoted by open 
symbols and the solid lines are fits to $dV/dI=R(T) \cosh(I/I_0)$. 
Temperatures from left to right are $4.12$, $3.92$, $3.80$, 
$3.64$, $3.45$, $3.36$, $3.26$, $3.16$, $3.07$, $2.80$ and $2.68$ 
K.} \label {fig:dvdi} 
\end {figure} 
 
\subsection {Determination of the linear resistance from high bias current  
measurements} 
\begin {figure} [t] 
\begin {center} 
\epsfig{file=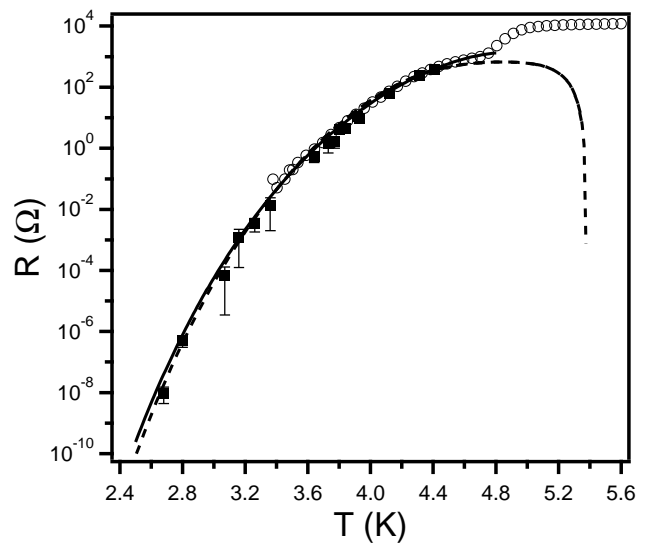, width=3.5in} 
\end {center} 
\caption {Resistance vs. temperature curve for sample B2. Open 
circles represent data that have been directly measured while 
filled boxes give the resistance values determined by fitting the 
$dV/dI$ curves of Fig. \ref{fig:dvdi} using the formula 
$dV/dI=R(T)\cosh(I/I_0)$.  The solid and the dashed curves give 
the best fits generated by the $R_{\text{WL}}(T)$ ($T_c = 4.81$ K) 
and $R_{\text{LAMH}}(T)$ ($T_c = 5.38$ K) formulas, respectively.} 
\label{fig:diff_res} 
\end {figure}

We now discuss the non-linear properties of films with constrictions. 
Measurements of the differential resistance versus bias current, 
$dV/dI$ vs. $I$, are plotted in Fig. \ref {fig:dvdi} on log-linear 
scale. Using these results it is possible to distinguish between 
the BKT mechanism, which leads to a power-law $V(I)$ dependence, 
and the phase slippage process, which is characterized by an 
exponential $V(I)$ dependence (Eqs. \eqref{eqn:lamh_voltage} and 
\eqref{eqn:lamh_dv_di}). From Fig. \ref {fig:dvdi} it is clear 
that at $T < T_{\text{BKT}}$ and sufficiently low currents the 
dependence of the differential resistance on bias current is 
exponential (it appears linear on the log-linear plots). Thus it 
is appropriate to compare the results with the LAMH theory. 
Equation \eqref{eqn:lamh_dv_di} can be written as $dV/dI = 
R(T)\cosh(I/I_0)$, where $R(T)$ is the temperature-dependent 
zero-bias resistance. Using this relation, we fit the differential 
resistance data and use $R(T)$ as a fitting parameter, as shown in 
Fig. \ref {fig:dvdi} by solid lines, each corresponding to a fixed 
temperature.\cite {i0} The fitting procedure illustrated in Fig. 
\ref {fig:dvdi} gives us a powerful indirect method of 
determination of the zero-bias resistance (it is implicitly assumed
that the ratio of the rates of thermally activated and quantum 
phase slips (if any) is independent of the bias current). This method 
is useful when the temperature is low and the resistance of the sample is 
below the resolution limit of the experimental setup. Thus, by 
fitting the $dV(I)/dI$ curves, we obtained the zero-bias 
resistance $R(T)$ down to very low values ($\sim 10^{-8}$ 
$\Omega$). This method was systematically applied on sample B2 and 
the results are shown in Fig. \ref{fig:diff_res} as solid squares. 
The open circles in Fig. \ref{fig:diff_res} represent the 
zero-bias resistance obtained by direct measurements at low bias 
currents. The two sets of data are consistent with each other. The 
solid curve in Fig. \ref{fig:diff_res} is a $R_{\text{WL}}$ fit 
obtained using Eq. \eqref{eqn:modified_lamh}. An excellent 
agreement is seen in a wide range of resistances spanning {\em 
eleven orders of magnitude}. This re-confirms that the thermally 
activated phase slip mechanism is dominant in the bridge  
samples\cite{nanowire_also} 
for $T<T_{\text{BKT}}$. We emphasize that the critical temperature 
of the bridge, which is used as an adjustable parameter, is found 
to be $T_c = 4.81$ K. As expected, the $T_c$ of the bridge is 
slightly lower than the critical temperature of the film 
electrodes $T_{c0} = 4.91$ K.   
 
The usual LAMH expression $R_{\text{LAMH}}$ (Eq. 
\eqref{eqn:lamh_ohmic}), which applies to thin superconducting 
wires, \cite{tinkham_lau,lau_etal,rogachev_bezryadin,bezryadin_etal} can 
also be used to fit our data. The overall shape of the fitting 
curve (dashed curve in Fig. \ref{fig:diff_res}) agrees with the 
data as well as with the $R_{\text{WL}}$ fit. The drawback of the 
usual LAMH formula is that the critical temperature of the bridge, 
which is used as an adjustable parameter, turns out considerably 
higher than the film transition temperature. 
For example, the dashed line fit in Fig. 
\ref{fig:diff_res} is generated using $T_c = 5.38$ K which is 
larger than the film critical temperature $T_{c0} = 4.91$ K. 
This apparent enhancement of the critical temperature of the 
bridge must be an artifact, 
because a reduction of the dimensions of MoGe samples always leads
to a reduction of the critical temperature.\cite {Oreg} 
On the other hand, the $T_c$ extracted from the fits made using Eq. 
\eqref{eqn:modified_lamh} are almost equal and slightly lower than 
the film $T_{c0}$ (Table I), as expected.  

A rapid decrease of the 
LAMH resistance at temperatures very close to the critical 
temperature reflects the behavior of the LAMH attempt frequency 
which approaches zero as $T \rightarrow T_c$. The LAMH resistance 
is proportional to the attempt frequency so we observe $R 
\rightarrow 0$ as $T \rightarrow T_c$ (dashed curve in Fig. 
\ref{fig:diff_res} ). Such behavior is unphysical and occurs since 
the LAMH theory is not applicable very near $T_c$. It should be 
emphasized that some of our measured bridges are wider than 
$\xi(0)$, yet the thermally activated phase slip model agrees well 
with the data. This is in agreement with the prediction (Ref. 
\onlinecite{langer_ambegaokar}, p. 510) that superconducting 
channels of width $w \lesssim 4.4\xi(T)$ should exhibit a 1D 
behavior, i.e. nucleation of vortices is unfavorable in such 
channels. Such condition is true for all of our samples. 
 
%%%%%%%%%%% 
% SUMMARY % 
%%%%%%%%%%% 
\section {Summary} 
Fluctuation effects in thin films interrupted by ``hyperbolic'' 
constrictions is studied. The measurements show two separate 
resistive transitions. The higher-temperature transition shows 
some properties of a BKT transition in the films (follows 
the HN formulae). The second apparent resistive transition is 
explained by a continuous reduction of the rate of thermally activated
phase slips with decreasing temperature. A quantitative 
description of the fluctuation resistance of narrow and short 
superconducting constrictions is achieved. For this purpose we have
modify the LAMH expression for the resistance of a one-dimensional nanowire. 
An indirect method that enables us to trace the resistance 
variation over eleven orders of magnitude is suggested, based 
on the analysis of the nonlinear effects occurring at 
high bias currents. The phase slippage model 
is found applicable in the entire range of measured resistances, 
suggesting that quantum phase slips\cite{lau_etal} do not occur 
in this samples, in the studied temperature interval, which extends
below $T_c/2$ for one sample (B1)). 
 
%%%%%%%%%%%%%%%%%%% 
% ACKNOWLEDGMENTS % 
%%%%%%%%%%%%%%%%%%% 
\begin {acknowledgments} 
We thank P. Goldbart and M. Fisher for suggestions.  This work was 
supported by the NSF carrier Grant No. DMR-01-34770, the Alfred P. Sloan 
Foundation, and the Center for Microanalysis of Materials (UIUC), which is 
partially supported by the U.S. Department of Energy Grant 
No. DEFG02-91-ER45439. S. L. C. thanks the support of NSF Grant 
No. PHY-0243675. 
\end {acknowledgments} 
 
%%%%%%%%%%%%%%%%%%%%%%%%%%%%%%% 
% BIBLIOGRAPHY AND REFERENCES % 
%%%%%%%%%%%%%%%%%%%%%%%%%%%%%%% 
 
%%%%%%%%%%%%%%%% 
% END DOCUMENT % 
%%%%%%%%%%%%%%%% 
\end {document}